\documentclass[sigconf, nonacm]{acmart}
\usepackage{hyperref}
\usepackage{url}
\usepackage{graphicx}
\usepackage{etoolbox}

\AtBeginDocument{%
  \providecommand\BibTeX{{%
    \normalfont B\kern-0.5em{\scshape i\kern-0.25em b}\kern-0.8em\TeX}}}

\copyrightyear{}
\acmYear{}
\acmDOI{}
\acmISBN{}

\makeatletter
\patchcmd{\maketitle}{\@copyrightspace}{}{}{}
\makeatother

\acmConference[]{}{}{}
 
\begin{document}

\title{CovidNet: To Bring Data Transparency in the Era of COVID-19}

\author{Kai~Shen}
\authornote{Corresponding authors with equal contributions}
\affiliation{%
	\institution{University of Kiel  \\
	Kiel, Germany}
}
	
\author{Tong~Yang}
\authornotemark[1]
\affiliation{%
	\institution{Boston College  \\
	Chestnut Hill, Massachusetts, USA}
}

\author{Sixuan~He}
\affiliation{%
	\institution{ADP  \\
	Roseland, New Jersey, USA}
}

\author{Enyu~Li}
\affiliation{%
	\institution{Texas Department of Transportation \\
	Austin, Texas, USA}
}

\author{Peter~Sun}
\affiliation{%
	\institution{Duke University\\
	Durham, North Carolina, USA}
}

\author{Pingying~Chen}
\affiliation{%
	\institution{1Point3Acres \\
	Las Vegas, Nevada, USA}
}

\author{Lin~Zuo}
\affiliation{%
	\institution{Duke University\\
	Durham, North Carolina, USA}
}

\author{Jiayue~Hu}
\affiliation{%
	\institution{New York University\\
	New York City, New York, USA}
}

\author{Yiwen~Mo}
\affiliation{%
	\institution{San Diego State University \\
	San Diego, California, USA}
}

\author{Weiwei~Zhang}
\affiliation{%
	\institution{South Dakota State University \\
	Brookings, South Dakota, USA}
}

\author{Haonan~Zhang}
\affiliation{%
	\institution{University of Texas at Dallas \\
	Dallas, Texas, USA}
}

\author{Jingxue~Chen}
\affiliation{%
	\institution{1Point3Acres \\
	Las Vegas, Nevada, USA}
}

\author{Yu~Guo}
\affiliation{%
	\institution{1Point3Acres \\
	Las Vegas, Nevada, USA}
}

\begin{abstract}
Timely, creditable, and fine-granular case information is vital for local communities and individual citizens to make rational and data-driven responses to the COVID-19 pandemic. This paper presents \textbf{CovidNet}, a COVID-19 tracking project associated with a large scale epidemic dataset, which was initiated by 1Point3Acres. To the best of our knowledge, the project is the only platform providing real-time global case information of more than \textcolor{black}{4,124} sub-divisions from over \textcolor{black}{27} countries worldwide with multi-language supports. The platform also offers interactive visualization tools to analyze the full historical case curves in each region. Initially launched as a voluntary project to bridge the data transparency gap in North America in January 2020, this project by far has become one of the major independent sources worldwide and has been consumed by many other tracking platforms \cite{dong2020interactive,wiki2020}. The accuracy and freshness of the dataset is a result of the painstaking efforts from our voluntary teamwork, crowd-sourcing channels, and automated data pipelines. As of May 18, 2020, the project website has been visited more than 200 million times and the CovidNet dataset has empowered over 522 institutions and organizations worldwide in policy-making and academic researches. All datasets are openly accessible for non-commercial purposes at \url{https://coronavirus.1point3acres.com} via a formal request through our APIs.
\end{abstract}

\begin{teaserfigure}
  \includegraphics[width=\textwidth]{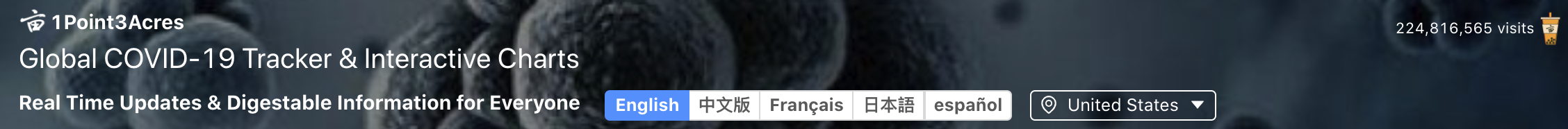}
  \vspace*{-5ex}
  \caption{Home page of 1Point3Acres CovidNet Project}
  \label{fig:teaser}
\end{teaserfigure}

\maketitle

\section{Introduction}
Starting from December 2019 or earlier, the outbreak initially detected and reported in Wuhan (Hubei, China) due to a novel type of coronavirus, the \emph{severe acute respiratory syndrome coronavirus 2} (SARS-CoV-2), has been rapidly spreading firstly across regions in China and other east-Asian countries, and then, since late February, to nearly all the continents in the world. As of \textcolor{black}{May 15}, there have been more than \textcolor{black}{4.42} million cases confirmed across 225 countries and regions, associated with \textcolor{black}{302} thousands deaths. Declared as a pandemic by World Health Organization on March 11, the coronavirus outbreak has brought severe challenges to not only local medical systems (especially in underdeveloped areas) but our society as a whole. 
\begin{figure}[htb!]
    \centering
     \includegraphics[width=0.47\textwidth]{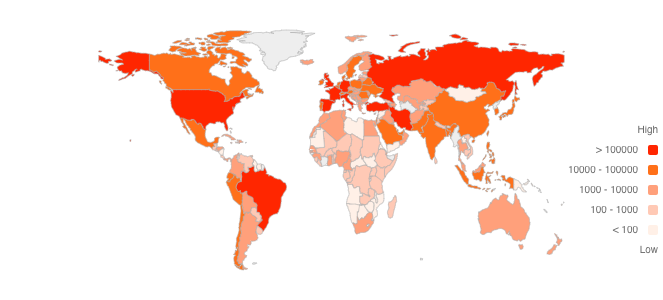}
     \includegraphics[width=0.47\textwidth]{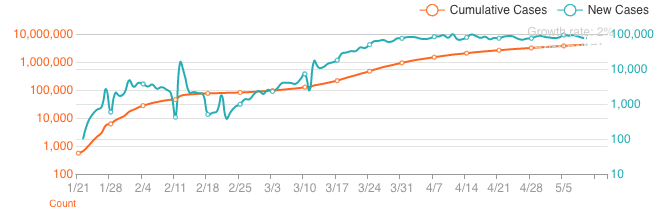}
    \caption{\small{Global COVID-19 case distribution and trend.}}
    \vspace*{-2ex}
\label{fig:global_distr}
\end{figure}

Governments across the world have been taking various measures at different levels in response to the pandemic. To be able to make scientific and data-driven decisions, local communities rely on timely and accurate epidemic data to understand the spread and the trend of the COVID-19 outbreak. Case information in cities and/or provinces is especially valuable to promptly adjust local policies in response to the rapid change of the pandemic situation. At the same time, nearly every individual's daily life has been severally affected by the crisis. To minimize the ramification of community spread, it is of extreme importance to provide the public with transparent and accurate local information to guide their daily life decisions. Overall, there is a huge need for \emph{timely, creditable, and fine-granular} data which can be easily accessed via a single platform.  
%
%
%

On January 31, our team initiated a project as a data website in response to the ongoing COVID-19 emergency. The original purpose of the project was to provide real-time case information in North America to compensate the delayed official reporting at the time. In less than three months, the platform has grown into a Global COVID-19 tracker which includes geographic information of COVID-19 spread with sub-division level breakdown for more than \textcolor{black}{27} countries and territories. Up to now, we have become one of the very few completely independent reporting sources of COVID-19 case data integration. As the United States has become the epidemic center of the world since late March where a real-time data collection across the country was unavailable from most official channels, we have been the \emph{original data provider} to the platforms like \cite{dong2020interactive, wiki2020} and the U.S. Centers for Disease Control and Prevention (CDC) since March 2020. To provide a full picture about COVID-19, we have also collaborated with another project team \cite{tracking} to integrate testing and hospitalization statistics in the U.S., and delivered a richer visualization of local pandemic status combined with our real-time case data. 
%

Our COVID-19 dataset, named as \textbf{CovidNet}, offers the full historical case trends with a fine-granular regional breakdown. The name addresses the hierarchical structure of the geographical network the data is embedded in.
CovidNet is constantly been updated to include the most up-to-date case information in real-time. To achieve data accuracy, real-time update, and worldwide coverage, the CovidNet features:
\begin{itemize}
    \item The data is collected from only reliable sources. Various quality control assurances have been applied;
    \item The data is updated in real-time with the effort of crowd-sourcing and automated data collection;
    \item The data is collected with fine geographical granularity worldwide.  
\end{itemize}
We have been providing data both to the public and academic institutes for pure research purposes. 
As of May 18, 2020, the project website has been visited more than 200 million times and has empowered over 522 institutions and organizations. We have built a convenient application programming interface to access the CovidNet dataset. And, to assist the worldwide battle against COVID-19, we would like to encourage more users, with both governmental and academic backgrounds, to take advantage of our data collection. 

The rest of the paper is organized as follows. Section~\ref{summarysection} provides an overview of the 1Point3Acres CovidNet project on both datasets and visualization. We discuss our data collection practice in details in Section~\ref{north_america} (North America data) and Section~\ref{worldwide} (global data). We elaborate our quality control mechanism in Section ~\ref{quality_control}, and introduce the rich set of interactive visualization tools in Section ~\ref{tools}. In Section ~\ref{access}, we explain briefly about how to access our datasets; related projects and platforms are mentioned in Section \ref{related}. We summarize our work in Section \ref{discussion} and discuss various ways that CovidNet could aid the battle against COVID-19. A disclaimer is highlighted in the end emphasizing the role of the CovidNet Project and associated contents.

\section{Project Overview}\label{summarysection}
Since the beginning of the COVID-19 outbreak, there has been overwhelming related information from numerous resources. A major challenge, therefore, is to integrate the scattered information on a single platform with consistent quality and credibility. The 1Point3Acres project focuses on the following three aspects in collecting and presenting the CovidNet data:
\begin{itemize}
    \item {\textbf{Data Accuracy and Consistency.}}
    We extract information from local health authorities and trustful media reports. Media reports are used when official data is significantly delayed, and are cross-checked with official data afterwords. No data from other tracking platforms is used in CovidNet to eliminate loops of references.
    \item {\textbf{Timely Update.}} 
    Since the launch of our project, CovidNet data has been updated in nearly real-time. Crowd-sourcing has been implemented to ensure timely updates. This sets us apart from official channels like WHO \cite{who_dashboard} and U.S. CDC whose updates are delayed by days. To our best knowledge, most other non-governmental platforms \cite{dong2020interactive, nyt} have also experienced a 1 - 2 days delay by far.
    \item {\textbf{Worldwide Data with Regional Breakdown.}}
    CovidNet provides case information with finer geographical granularity in over \textcolor{black}{27} countries and we are still expanding the coverage. After the initial launch with county-level case data in the U.S. and Canada, we have received numerous feedback from local authorities and residents on how it helped local communities in decision making, which motivated us to bring the finer data granularity to more countries impacted by COVID-19.  
\end{itemize}

\subsection{CovidNet Dataset}
\label{dataset_summary}
Our CovidNet dataset provides real-time epidemic information in three major categories: confirmed, deceased, and recovered cases when they are publicly available\footnote{We follow the definition of different categories provided by health authorities in each country.}. The complete history since the outbreak in each local region is available in the CovidNet dataset. Table~\ref{tab:summary} summarizes granularity, information source, and update frequency of the current dataset\footnote{The geographical granularity differs in different countries, although one shall \textit{not} interpreter that the official US/Canada data has a better granularity than ones in other countries. While city level and per-case level case data are accessible in many regions outside U.S./Canada, systematically aggregating such information worldwide, however, is beyond the capability of our voluntary team.}. 
\begin{table}[!hbt]
    \centering
    \begin{tabular}{|p{2.4cm}|p{1.7cm}|p{2.1cm}|p{0.9cm}|}
    \hline
    & U.S./Canada & Hardest Hits* & Others  \\ \hline
    Granularity  & county/city & province/state & country \\ \hline
    Number of regions & 3169 & 974 & 161 \\ \hline
    Original sources & health dept. \& local media & health dept. \& state institution & WHO \\ \hline
    Update frequency & 1 hr & 1 hr & 2 hrs \\ \hline
    Number of source  & 1064 & 974 & 1 \\ \hline
    Cumulative cases & \textcolor{black}{1.5M} & \textcolor{black}{1.9M} & \textcolor{black}{1M} \\ \hline
    Total Deaths & 92K & 161K & 49K \\ \hline
    \end{tabular}
    \caption{Summary of CovidNet dataset. *We now have included \textcolor{black}{27} countries, with more to come. }
    \label{tab:summary}
\end{table}
\subsection{Interactive Tools on 1Point3Acres Tracker} 
\begin{figure}[!htb]
  \centering
  \begin{minipage}[b]{0.30\textwidth}
    \includegraphics[width=0.75\textwidth]{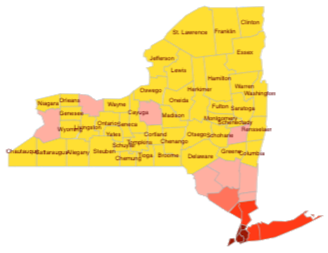}
  \end{minipage}
  \hfill
  \begin{minipage}[b]{0.15\textwidth}
    \includegraphics[width=\textwidth]{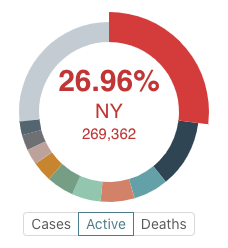}
  \end{minipage}
    \vspace*{-3ex}
    \caption{\small{Example visualization tools for geographical distribution. Left: county level interactive case map in state of New York, U.S. Right: a doughnut chart of case distribution in U.S by states. Figure best visualized in color.}}
\label{fig:geo_summary}
\end{figure}
\begin{figure}[htb!]
    \centering
     \includegraphics[width=0.45\textwidth]{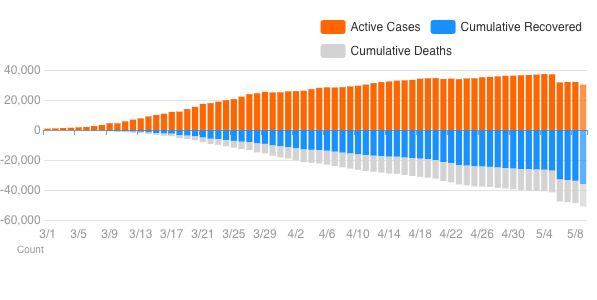}
     \vspace*{-3ex}
    \caption{\small{Example burn down charts for visualizing trends of active, deceased, and recovered cases in Lombardia, Italy.}}
\label{fig:trend_summary}
\end{figure}
\noindent With a flood of available COVID-19 data, we offer a suite of interactive analysis and visualization tools to provide the general public more insights about the current situation of the COVID-19 pandemic. We focus on presenting both the \textbf{temporal trends} (i.e., epidemic curves) as well as the \textbf{geographical distribution} of the case spread. (See Section \ref{tools} for details.) The interactive tools are available in four different languages. 

\begin{itemize}
    \item {\textbf{Geographical Distribution.}} The COVID-19 case distribution is presented in various ways, including \textit{epidemic maps}, \textit{doughnut charts} and \textit{tabular views}. Figure~\ref{fig:geo_summary} showcases a few examples of visualization. World, State, and county level epidemic maps are provided for user exploration. The tabular view can be customized to rank by different case dimensions, such as the infection rate, the death rate, etc. 
    \item {\textbf{Temporal Epidemic Curves.}} Temporal trends of the outbreak are captured in various epidemic curves, including basic \textit{time-series lines} and \textit{burn down charts}. Users can choose and compare curves of different regions with our interactive tools. One such example is shown in Figure~\ref{fig:trend_summary}. 
\end{itemize}
\section{North America data collection}\label{north_america}

The United States has been the epidemic center since March, 2020, and the most challenging area to integrate creditable case information in real-time due to data inconsistency at different levels of the public health system. 
Our platform was initially launched to confront this challenge and the severe delay by official authorities. This section describes our data collection and validation practice during different stages of the outbreak amid a timely, reliable, and county-level COVID-19 dataset. 

Before discussing the data collection practice in detail, we highlight the specific features of the CovidNet dataset in the U.S. and Canada in addition to Section~\ref{dataset_summary}. The tracking in North America started on January 21, 2020 when the first case was officially confirmed in the U.S. 
The data covers \textcolor{black}{3169} sub-country-level regions across the North America\footnote{This includes all counties/districts in U.S./Canada, U.S. Federal Bureau of Prisons system, U.S. Military and Veteran Affairs systems, and U.S. territories. Cases from cruises have been separately presented.}.
All data is collected from publicly available sources, including both local health official announcements and reliable media reports, and is integrated from \textcolor{black}{1064} distinct websites. In addition to case information in the CovidNet, our project also includes testing locations and statistics thanks to the COVID Tracking Project \cite{tracking}. 
\subsection{A Change Log of Data Collection Practice}\label{datac}
The epidemic situation underwent several stages in North America since January 2020, each with unique challenges in data collection and validation. We now elaborate different stages of the data collection practice along with the evolution of the COVID-19 outbreak.

\subsubsection{\textbf{Initial stage: January to Late February}}\label{datac_1}
The first North America COVID-19 case was officially reported on Jan 21 \cite{holshue2020first} and the total cases had remained at a low level until late February\footnote{As of Feb 29, 2020, a total of 72 cases were officially reported in U.S. and 20 in Canada.}. During this period, both federal and local official health departments had not developed a systematic reporting schedule. Each individual case, however, received plenty of media coverage with both geographic and demographic details. The most timely data source had been local media reports. 

The CovidNet project was initially launched in Jan 31, 2020 as a platform for real-time case data aggregation of related media report\footnote{Our searching had traced media sources including \emph{CBSN (U.S.), Seattle Times (U.S.), Global News (Canada), National Post (Canada)}, etc.}. Since active crowed-sourcing was the most effective way to track the sparsely emerging cases, we organized a volunteer team working 24/7 for real time updates and focused on data accuracy. 
We recorded each single confirmed case with all accessible data features including geo-location, demographic information, infection cause, and a summary of news report. In additional to per-case information, the reference links to original news were also attached to each record in our dataset and were visible to all users on the website. This helped us account for double counting and to cross validate against later official reports. 
\subsubsection{\textbf{Expanding stage: Late February to End of March}}\label{datac_2} 
The epidemic outbreak in North America started expanding geographically since late February. While numbers were still closely tracked by local media, the task of integrating all available sources manually by volunteers had become more and more challenging. Health official departments also started to actively release case information but yet to provide a real-time data.
With limited capacity, in addition to active searching for media reports by volunteers, we 
offer an issue report submission form to all users, through which anyone can provide useful information back to us. The submission form categorizes information reports into 9 different classes: \emph{"New Case", "Recover Case", "Death Case", "Error Report", "Feature Request", "Breaking News", "Further Details", "Testing Location",} and \emph{"Question"}. Whenever a delay of our data or an error was noticed, a user could directly report the issue to us with trustful links. The reputation we built up during the initial stage has rewarded us with a large number of visits and users, which in turn provides us informative feedback frequently. As of \textcolor{black}{May 14, 2020}, we have received and resolved \textcolor{black}{16240} issue reports from our users, which have been a significant part of the data source. After an issue gets reported by users, a volunteer would be assigned to manually check the provided information (via media or official sites), and compare the report with all currently recorded cases (within the same county). 
\footnote{We have also used the help from several twitter alarm accounts at the beginning of this stage. }

\subsubsection{\textbf{Rapid Increasing Stage: Since April}}\label{datac_3}
The total number in the U.S. surpassed 10,000 on March 27, 2020, which significantly elevated the difficulty for data collection and validation. While local media remained to be a valuable data source, in regions with large increasing numbers, there appeared to be also a delay even in media reports. On the other hand, more local health departments started to develop official announcement schedules (generally from 1 to 3 times per day). Therefore, while still accumulating local media reported data in real-time, we attempted to adapt our collection process to include most up-to-date official health reports, which would prevent a possible delay in the data integration procedure. We now give a brief introduction about the construction of this pipeline by using U.S. health systems as an example. 

There is a three-level hierarchy structure of the official public health system: the overall country-level CDC, the state-level health departments, and local county-level health departments. With a bottom-to-up data flow, the CDC database has experienced the most delay, which, is also the exact reason to launch a project like ours. As for state-level and county-level reports, the situation has been mixed: some state health departments have tracked local cases closely and hence provided a trustful source for update checking, while others are in general behind county-level statistics. There are in total 3,243 counties in the United States.
To monitor all local reports manually is impossible for the volunteer team. We, therefore, initiated an automated data fetching pipeline, which enables a 24/7 checking for updates from local health department sites. 
Any fetched data update would be assigned to the volunteer team for a second-round check, as official reports potentially contain several types of noise which we discuss in detail in Section 5.  
Only verified updates would be finally recorded in the database. 

We want to address the three important facts below about our automated official-data checking pipeline:
\begin{enumerate}
	\item We collect both official health authority reports and media report to stay close to real-time. 
	\item While most official health authority provide case data to public, it remains of significant importance to integrate all local data into a complete dataset. 
	\item We account for potential noise in data from local health officials and implement various quality assurance steps before adding into our database. 
\end{enumerate}
This automation pipeline has been integrated into our workflow since April, 2020, and has become one of the dominant components of our data collection process.
\subsection{The Structural Evolution of the Data Format}\label{datafmt}
We  used the cloud collaboration service provided by Airtable. Our team members can simultaneously work on the same tabular data. There are three major tabular formats we have used:
\begin{itemize} 
	\item \textbf{Expanded Tabular (ET) format: } we document every case (sometimes a small cluster of cases) into a single row to provide a per-case view to our user. We also provide a summary of the original report for each case-cluster. In this ET format, all information including the reference link to the source is presented directly to users for the full transparency. 
	\item \textbf{Compact Tabular (CT) format: } given a table content topic (e.g. recovered case table) and a fixed (geographic) granularity, we assign each row  with a region, and each column with a date. The number in a cell therefore indicates the accumulative counting in the row-labeled region on the column-labeled date. And each row represents a time series of data in a specific region. While CT is easy to consume compared to ET, the CT format cannot easily summarize the full set of references and losses certain demographics information. We keep such dimensions only internally visible for quality control purpose.
	\item \textbf{Statistic Assistant (SA) format: } this type of table is used to present statistics of currently collected data. We usually assign each row a region, which could be of country-equivalence level, state-equivalence level, or county-equivalence level. Different columns  are used to represent different information categories associated with each region like \emph{region name, total confirmed number, fatality rate,  contact of the local health official}. SA is mostly used for tabulated presentation and other interactive maps on our project web-page. 
\end{itemize}
%
%
During initial stage (section 3.1.1) , we used a single database to record all cases in the ET format. It was later divided into two separate ones in early March: one for U.S. cases and the other for Canada cases. Corresponding SA tables have also been updated automatically since then. 

The ET format continued to be the dominant one until the end of the Expanding stage (section 3.1.2). Maintaining ET format when U.S. official only reports aggregated statistics became impossible and we have transformed our data table into the CT format in late March. 
%
We kept deceased case table and recovered case table in ET format until mid-April and transformed into CT format then. 

\section{World Wide Data Collection}\label{worldwide}
Most mainstream COVID-19 data platforms \cite{wiki2020, dong2020interactive, 1P3A, who_dashboard, worldmeter, DXY, nyt} offer case information at country-level, with only exceptions for China by \cite{DXY} and North America (e.g. by our project and subsequent platforms that aggregate our data \cite{dong2020interactive, wiki2020}). International organizations like WHO \cite{who_dashboard} and ECDC aggregate worldwide data but are often delayed without sub-division regional breakdown. Creating a platform with consistent freshness, credibility, and granularity for worldwide data is the key to present the full picture of the COVID-19 spread, which, however, poses a great challenge. 

Determined to confront this challenge, we have exploited a combination of manual data collection, crowd-sourcing channels, and automated data pipelines to solve the problem. As of \textcolor{black}{May 18, 2020}, the CovidNet dataset has covered  data in \textcolor{black}{1055} state-level regions and \textcolor{black}{3169} county-level regions of more than \textcolor{black}{27} countries and territories. While collecting the most updated data in a timely fashion, we also retrieved all the case history since the outbreak within each region.

We choose to only present daily aggregated numbers in CovidNet in each region for consistency. We note that some health departments have provided richer information than a daily aggregated number. A full list of countries can be found in Appendix. We prioritize countries with the most confirmed cases and attempt to cover major countries in all continents of the world. We highlight that this is an ongoing effort to expand to as many countries as our resource permits.

\noindent\textbf{Subdivision Naming Convention.} We adopt the ISO-3166 standard \footnote{\url{https://www.iso.org/iso-3166-country-codes.html\#2012_iso3166-2}} to normalize provinces or states names in CovidNet. Official  names in English and local languages are both available.

\subsection{Case Reporting Paradigms in Different Countries}
\label{reporting}
Local health authorities reports are used as data source for consistent credibility\footnote{The only exception is North America discussed in Section~\ref{north_america}.}. While the data reporting systems differ across countries, they mostly fall into the following two categories. 

The first class of countries offers open access to the full historical data starting from the outbreak. Within this paradigm, case information is provided in either the per-patient level or per-region daily statistic level. Examples are the public repository of Italy \footnote{Presidenza del Consiglio dei Ministri  official github repository: \url{https://github.com/pcm-dpc/COVID-19}} and the open data API of Colombia \footnote{El portal web de Datos Abiertos del Ministerio de Tecnologías de la Información y las Comunicaciones: \url{https://www.datos.gov.co/en/Salud-y-Protecci-n-Social/Casos-positivos-de-COVID-19-en-Colombia/gt2j-8ykr/data}}.  We transform all data into \emph{per-region-day} statistics and discard detailed demographics. While this simplifies the full history retrieval, it requires systematical quality control, due to the modification of historical data and changes of reporting criteria. For instance, the official data repository of Spain changed confirmed case definition and data fields in late April, which led to mismatches in our time-series within a short time window. \footnote{ \url{https://cnecovid.isciii.es/covid19/\#documentaci\%C3\%B3n-y-datos}} We apply alarms whenever a suspicious decrease in the curve gets observed, which is then assigned to manual check by volunteers.

The second class of countries reports only the most recent case data, through a data-accessing interface. In most locales, the historical data is collectively archived and stored in various formats (pdf, csv, json), or retrievable from official daily reports. Examples include the official COVID report by Korean CDC\footnote{Press Release Archive of Korean Center of Disease Control: \url{https://www.cdc.go.kr/board/board.es?mid=a30402000000&bid=0030}} and the South Africa COVID news portal\footnote{\url{https://sacoronavirus.co.za/category/press-releases-and-notices/}}. We collect all archived information and create an integrated dataset. As in such data-accessing systems, there lacks a consistent way to track edits in historical data, we hence only retrieve the historical data once. 
Occasionally, we failed to trace back all the historical data archive. We resort to local non-official data platforms to back fill the earlier case data\footnote{We specifically acknowledge the effort from \cite{de_bruin_j_2020_3711575} for their endeavor.}.

\subsection{Data Collection Pipeline}\label{world_pipline}
We check all the official data sources every 2 hours and update the most recent data in our database accordingly. For countries that provide open access to the full case history, we also implement scheduled (daily) checking for modifications of official historical data. The timestamp associated with each record is in agreement with the local time of the publishing authority\footnote{We notice that a subset of countries provides official data with a 1-day delay, which is annotated accordingly on the web page.}.

Given different stages of the outbreak, the data accessibility and publishing channels are constantly evolving in each country. For instance, official data is provided in per-patient level in the initial stage in most countries, and slowly evolves into an aggregated format during the outbreak stage. We combine our data pipeline along with the volunteer's manual effort to ensure the data quality and consistency. Section~\ref{north_america} highlights this issue in the North America as an example.

\section{Data Quality Control}\label{quality_control}

The data quality control (QC) is always the first priority of the CovidNet project, as both local communities and researchers might make critical decisions based on the data. Note that we only have access to the case information released by health officials and/or media reports, and the quality of the original case data is beyond our capability. We control the CovidNet dataset to accurately reflect the information provided by local health authorities. This section focuses on specific quality control challenges in a rapidly evolving pandemic and the practices we adopted in the voluntary teamwork.

\subsection{Quality Control with different Reporting Paradigms Worldwide}
As discussed in Section~\ref{reporting}, our data comes from distinct reporting systems in different countries. While each paradigm requires a specific quality model, the general principle is to start with the data with the finest granularity and/or with the full history. For instance, when per-case data are provided (e.g. open data projects in Columbia \footnote{El portal web de Datos Abiertos del Ministerio de Tecnologías de la Información y las Comunicaciones: \url{https://www.datos.gov.co/en/Salud-y-Protecci-n-Social/Casos-positivos-de-COVID-19-en-Colombia/gt2j-8ykr/data}} and Philippine \footnote{ Department of Health Data Drop \url{https://ncovtracker.doh.gov.ph/} } health ministries), the pipeline would aggregate from such per-case dataset to get daily case statistics and geographical distributions. Following the same principle, when health officials release and update the entire aggregated daily history (e.g., Italy and Spain), we update the entire time series whenever edits are made officially. 

We dedicated our effort mostly in health reporting systems with a federated administration structure. The key challenge presented with such a system is the \textit{asynchronized} natural of data from different levels. Taking the United State as an example: county health departments release case statistics following their own schedules while state officials may update at a lower frequency, typically 1-3 times a day. At any given time, aggregated county-level data may differ from state-level ones. In extreme cases, state-level reports could be delayed by days. This asynchronized nature has brought difficulty for QC. The situation gets further complicated when an \textit{unassigned} category is presented. 

\textbf{Quality Control practice in federated systems}: As discussed in section~\ref{north_america}, our database records the finest granularity, and has included the full set of references either publicly visible (in the early stage) or internally traceable (in the current stage). All volunteers follow the same protocol to determine whether an inconsistency is due to human mistakes or not, by investigating data from different levels of health departments and also internal comments. On the website, a note would be left to users when presented numbers differ from official statistics. We also created an \textbf{Inconsistency Diary} (maintained by the volunteer team) as an internal reference to track discrepancies, and periodically revisit issues that persist.

\subsection{Timely Updates with Potential Noise in Official Data}
We acknowledge that any case data is subject to edit and change in a rapidly growing pandemic situation, and have assumed potential noise in the official data collected. While most of noise would be corrected eventually, to present timely update, we have categorized several noise mechanisms and implemented quality assurance accordingly. 

\begin{itemize}
    \item \textit{Suspicious jump in case number:} an issue would be created when a large jump in case numbers appears\footnote{The definition of jump is determined empirically in practice. For example, a number change larger than 3 times would be defined as a jump, given that the initial value is larger than 100.}, and updates in the sub-region would be temporarily suspended (typically for 2 to 6 hours) to allow for potential subsequent corrections in official data. 
    \item \textit{Decrease in accumulated numbers:} most number decreases would be considered as history corrections\footnote{A large decrease might be an error, and falls into the \emph{jump} category described earlier.}. When this happens, if the full official historical data is accessible, we update the dataset according to the most recent official time series. In systems where only the current statistics is available, we adjust the most recent historical case data to maintain the non-decreasing property of each time series\footnote{A longer time-series may requires modification in practice. This may not be optimal but the ideal solution requires data we have no access to.}. 
\end{itemize}
In a nutshell, continuous update mechanism is instrumental in reducing noise in the CovidNet dataset. We also implemented daily-scheduled checks and cross validations with official and other independent platforms. Many of the noises have been identified and corrected thanks to the manual efforts of our volunteer team. 
\subsection{Quality Control in a Decentralized Volunteer Team }
Unlike many other platforms, the CovidNet project has been conducted by a fully decentralized volunteer team. This presents a unique challenge to the data control quality. 

\noindent\textbf{Repeated entry of the same data}\label{errhuman_1}
In the decentralized team, multiple volunteers might be working on the same update (\emph{overlapping work}), or failing to recognize existing duplicated records (\emph{duplicated records}). 

To prevent overlapping work, our volunteers will first check potentially related open updates before their own updates. 
Duplicated records was prominent in the early stage of CovidNet project when we used ET format. If accumulated numbers were available in the reports, the volunteer would directly check the total number in the specific area, and prevent duplicated cases. If accumulated numbers were unavailable, the volunteer would check within all existing records, especially case-clusters sharing the similar attributes as described in the report, and leave out duplicated reports.

\noindent\textbf{Uncoordinated Deployment}
All volunteers can deploy data updates to the front-end website. This was designed earlier to avoid delayed data presentation to the public. However, when one attempts to deploy a correctly completed number change, other members might be in the middle of updating and validating case information in a different region. 
For instance, on April 15, 2020, when a volunteer attempted to modify the confirmed number in Okaloosa, Texas from $102$ to $103$, the data entry experienced a transit state as "$102103$", which was quickly corrected by the volunteer. However, during the short transit stage, there happened to be a website deployment, which resulted in an abrupt number jump by more than $100000$, and then got corrected within minutes after the subsequent deployment. \footnote{This accidental jump then appeared rapidly also on other platforms \cite{dong2020interactive, worldmeter} and media reports.}


These type of issues are not typical "data errors", but due to the decentralized deployment setup, which is, however, necessary to prevent delayed website presentations. To minimize disadvantages associated with, we gradually constructed a list of rules that would be automatically checked before each deployment. Any deployment would be forbidden (unless being manually interfered), if one of the following criteria is triggered: 
\begin{enumerate}
	\item the number is smaller than the one from the previous day;
	\item the number of a county-level region increases more than $4000$ in a single day;
	\item the daily increase is more than $300\%$, while the previous day's number is larger than $10$;
	\item the daily increase is more than $200\%$, while the previous day's number is larger than $50$;
	\item the daily increase is more than $50\%$, while the previous day's number is larger than $1000$;
\end{enumerate}
where all thresholds applied above are deduced empirically.
\section{Interactive Tools for visualization}\label{tools}
%
The CovidNet project offers a rich set of data visualization tools for all users visiting our website. As outlined in Section~\ref{summarysection}, these interacted tools are integrated for better presenting both the temporal trends and the geographical distribution about the COVID-19 pandemic. We discuss the details in this section. 
%
%
\subsection{Geographical Distribution of Current Cases}
\label{geodistribution}
The geographical distribution of the COVID-19 spread is important for communities and individuals to make decisions. 
We visualize the current case statistics via two visualization modules: \emph{epidemic maps} and \emph{doughnut charts} as showcased in Figure~\ref{fig:geo_summary}. For both the whole world and each single country, we present a detailed epidemic map which shows the statistics\footnote{Statistics includes local populations, confirmed numbers and density (per million), deceased numbers, deceased density (per million population), testing numbers (if available),  and testing density (per million population).} in each state-level region, where darker colors indicate a larger number in the area. We also implement doughnut chart representation mostly to spot areas associated with largest proportion of the cases.
%
\subsection{Temporal Epidemic Curves}
\label{epicurve}
The time series of case data, often referred to as epidemic curve, helps general public to understand the full history and the current stage of the COVID-19 spread. Our project includes both basic trends charts with plain numbers and burn-down charts which provide further insights into the varying condition. Users can explore these epidemic curves for all sub-divisions in our dataset. 
%

\noindent{\textbf{Basic trend charts}} We provide basic trend charts at daily resolution for the following type of statistics: confirmed cases, deceased cases, positive/negative testing cases, positive testing rates, and hospitalized cases.
%
For both confirmed and deceased curves, users can choose to view them in either linear or logarithm scale. 
For example, Figure~\ref{fig:global_distr} shows the cumulative and daily new confirmed cases worldwide. 
%

\noindent{\textbf{Analytic trend charts}}
We have implemented the \emph{burn down chart} to show simultaneously the trends of active cases and recovered/ deceased cases, which provides a clearer picture about the local progress over time (see Figure~\ref{fig:trend_summary} ).

\subsection{Cross Region Comparison}
\begin{figure}[ht!]
    \centering
    \includegraphics[width=0.45\textwidth]{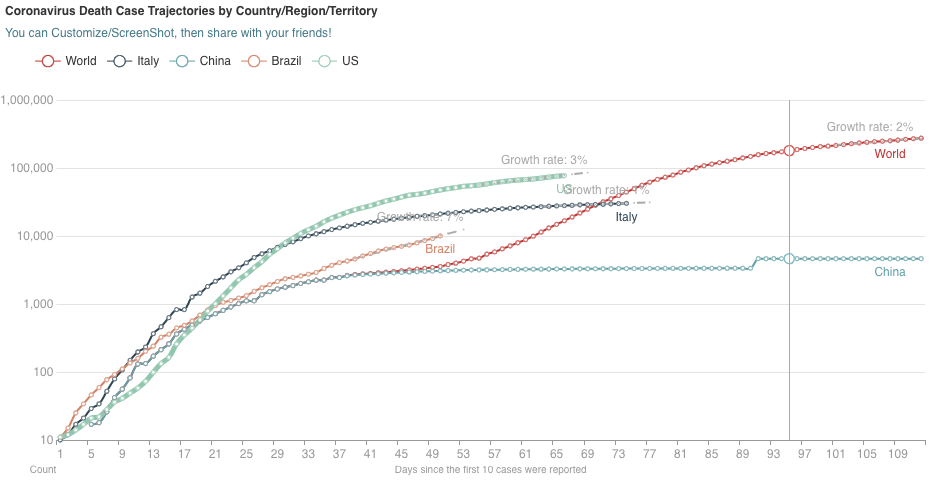}
    \vspace*{-3ex}
    \caption{Comparison of the total confirmed cases in a few countries. X-axis are the relative days since the case reached 100 in each location.}
\label{fig:log_trend}
\end{figure}
We offer several ways for users to compare the COVID-19 trends across different countries and sub-divisions over the whole period. One such example were shown in Figure~\ref{fig:log_trend}. User can choose the set of countries or sub-divisions they are interested in. In addition to static plots, we make available several animations on our project website to show the evolution of the comparative statistics across different regions. 
\section{Accessing CovidNet Dataset}\label{access}
We would grant open access to CovidNet for any non-commercial data usage. Our dataset is accessible by filling a data request form at \url{https://airtable.com/shrMqS4C6wjpZLCP0}, before which the user should have read carefully about terms of data usage at \url{https://coronavirus.1point3acres.com/en/data}. We do prohibit crawling, scraping, caching or otherwise accessing any content on the platform via any automated means, due to the expensive bandwidth consumption which has brought a huge financial burden on the project. To continue providing timely information to the general public, we welcome only fair access to our dataset and other contents.

\section{related work}\label{related}
We briefly discuss some other platforms which are also taking efforts in serving COVID-19 related epidemic information to the general public. 
\subsubsection*{Dingxiangyuan \cite{DXY} } 
Initiated in early January, 2020, the Dingxiangyuan (DXY) project has been among the most popular dashboards for COVID-19 information in China. They provide daily update by collecting data from all levels of official health departments, which has been quite trustful, and we have adopted their data as our source in China. Compared with the case in U.S., the data collection in China has been relatively easy and well-organized, as official channels have started a formal reporting schedule since the early stage and no extra media reports were required. We have seen the success and the impact brought by DXY's integrated data platform, which helped local communities in China to response effectively to the outbreak. This inspired us to start our own project for North America and eventually as a global tracker. We also studied their data presentations which helped us build our own system. While their focus has been the condition in China, data in the rest part of the world is collected and presented only in country level, with a delayed update for data in U.S. due to the reason we explained earlier.

\subsubsection*{John Hopkins COVID-19 Dashboard \cite{dong2020interactive}  }
A previous paper\cite{dong2020interactive} described the effort of a team from John Hopkins University in a global COVID-19 data platform which consumes dataset provided by DXY in China and later ours in U.S for a global visualization. As a completely independent data-collection source, our North America dataset has been continuously used by \cite{dong2020interactive} as one of their sources in the county-level breakdown data in the U.S. area. Unlike our close to real-time update, there has been a delay of U.S. data on the JHU Dashboard (1 or more days) since the end of April\footnote{This had urged us to enhance our own data quality to prevent errors when serving as the most up-to-date independent data platform in U.S. to the public.}. Last but not least, similar as DXY, worldwide data is only provided at country level on the JHU Dashboard. 
%
\subsubsection*{Worldmeter COVID-19 Tracker \cite{worldmeter} }
This is another COVID-19 tracker which offers partial regional breakdown in U.S. while providing country-level numbers in the rest of the world. For several states in U.S.\footnote{As of May  8, 2020, the county-level breakdown is provided for 8 states in U.S. by \cite{worldmeter}, including NJ, CA, PA, FL, TX, LA, OH, and WA.}, Worldmeter updates information in real-time which has been more up-to-date than the JHU Dashboard. 
\subsubsection*{Other related projects}
The COVID Tracking Project \cite{tracking} has collected and provided data in testing, hospitalization, and, very recently, the demographic distribution in different states in U.S.. We have been using their data for both testing and hospitalization visualizations, which provide an enriched description about local epidemic conditions especially when the testing does not widely cover the local community.
The DriveThruLocation \cite{tracking} is another valuable project which collects detailed information on locations providing drive-through COVID-19 testing services in U.S.. While not closely related to the pure data practice, we have collaborated with this project team as testing information is vital for local communities, which fits our purpose of serving the public in general.

\section{Discussion}\label{discussion}
This paper delivers a detailed introduction about the 1Point3Acres CovidNet project, elaborating both the data collection process and the quality control mechanism.
 
We would like to share some lessons we have learnt from the project.
The real time nature and the exhaustive geographical distribution of CovidNet have attracted a large number of users on the platform. 
While we have been trying our best to provide a creditable data integration, our practice has also suggested the necessity of constructing an official information integration pipeline to confront potential public health challenges, especially in the current era when human interaction has become an essential component of the modern society. 
At the same time, as inter-national connections have been much stronger than any previous time, global level pandemic information sharing would be of vital importance in both learning from experiences of other countries and assessing a local public health risk level according to interactions among different countries and regions. 
On the other hand, when more public challenges have upgraded to the global level, we have also seen the strength of data driven approaches in tackling large scale problems. 

With real time update of sub-division level COVID-19 information across more than \textcolor{black}{27} countries, the CovidNet dataset could benefit everyone in various ways. 
For general public, the analytic trend charts have provided more intuitive descriptions about the local epidemic situation, and would help local communities to make decisions about working and traveling.
For local governors, a comparison with situations and trends in other states/counties would be instructive for evaluating different policy scenarios, including restrictive orders and economy reopening.
For academic community, the CovidNet dataset could lead to researches in diverse potential directions: 
time series data could be combined with conventional epidemiological models to make prediction about the near future; 
the sub-division breakdown has provided detailed geographical distribution of COVID-19 outbreaks and may be used to analyze the impact of different external factors associated with each local region, e.g. weather, economy, population, industries, races, restrictive-order levels, and so on;
the worldwide dataset together with information on inter-national activities offers the potential to study the global spreading behavior of the disease.

We welcome and encourage users from all industries utilizing the CovidNet to assist the battle against COVID-19.
We especially look forward to more insights offered from academic researches by investigating the rich dataset provided by CovidNet project.

\section*{Disclaimer}
The CovidNet project, including all data, mapping, analysis, copyright 2020 1Point3Acres, LLC, all rights reserved, is provided for the public with general information purpose only. All information is collected from multiple publicly available sources that do not always agree. While we will try our best to keep the information up to date and correct, we make no representations or warranties of any kind, express or implied, about the completeness, accuracy, reliability, with respect to the website or the information. We do not bear any legal responsibility for any consequence caused by the use of the information provided. We strictly prohibit unauthorized use of the information in commerce or reliance on the information for medical guidance. 1Point3Acres disclaims any and all representations or warranties with respect to the project, including accuracy, fitness of use, and merchantability. Screenshots of the website are permissible so long as appropriate credit is provided.

\section*{Acknowledgement}
We would like to thank all contributors to the CovidNet projects who have not been directly listed in authors, including all engineers and volunteers: Yixin Wan, Maggie Hou, Lily Wang, Daisy Fang, Mark Lee, Yao Xu, Vivian Jiang, Chenyang Wu, Mingtian Zhou, Zhenzhuo Lan, Hongmin Li, Yun Han, Jie Zong, Clara Zhang, Gretchen Zhang, Huijie Tao, Huihui Shang, Zhenyu Jiang, Minchen Wang, Chuya Guo, Krystal Zhang, Konka Shi, Jonathon Jing, Xuan Chen, Ziyi Yan, Junrui Zhao, Yi Chen, Allison Li, Liyan Tian, Jiuyue Cai, Ning Cao. 
Besides, we want to thank all our users, who have continued provided us with data source associated with excellent advice, which helps us keep improving the project. 

At the same time, the whole project would not be possible without help from other teams and organizations.
We would like to thank other project teams who have shared rich data and information to us, including The COVID Tracking Project and the DriveThruLocation Project. 
We also want to give special thanks to Airtable, who granted us a package with 5000 free credits for the usage of upgraded plan which make the collaboration possible with a large volunteer team. We claim no conflict of interests.

\bibliographystyle{ACM-Reference-Format}
\bibliography{main}

\appendix
\begin{table*}
\centering
\caption{List of official worldwide data source}
\begin{tabular}{|p{2.4cm}|p{3.7cm}|p{4.4cm}|p{4cm}|}
\hline
\textbf{Country} & \textbf{subdivision} & \textbf{Information Source} &  \textbf{Available Data} \\ \hline
Austria & Bundesland & \href{https://www.sozialministerium.at/Informationen-zum-Coronavirus/Neuartiges-Coronavirus-(2019-nCov).html}{Bundesministerium für Soziales, Gesundheit, Pflege und Konsumentenschutz} &  confirmed, deceased, recovered  \\ \hline
Belgium & Region/ Province & Sciensano, the Belgian Institute for Health & confirmed, deceased, recovered  \\ \hline
Switzerland & Conton (States) & Zurich Statistisches Amt & confirmed, deceased, recovered \\ \hline
Chile & Region (Province) & Ministerio de Salud &  recovered at country level \\ \hline
Germany & Stadtstaaten/ Flächenländer  & Robert Koch Institute  & recovered info not available \\ \hline
Spain & Autonomous Communities  &Instituto de Salud Carlos III   & confirmed, deceased, recovered \\ \hline
United Kingdom & Upper Tier Local Authorities & National Health Service    & confirmed, deceased, recovered \\ \hline
India & States/ Territories & Ministry of Health &  confirmed, deceased, recovered \\ \hline
Italy & Region & Presidenza del Consiglio dei Ministri &  confirmed, deceased, recovered  \\ \hline
Japan & To/ Dō/ Ken/ Fu & Ministry of Health, Labour and Welfare  & confirmed, deceased, recovered \\ \hline
South Korean & Do/ Jachido & Korean CDC, \& Ministry of Health and Welfare & confirmed, deceased, recovered \\ \hline
Malaysia & States/ Territories & Director General of Health Malaysia &   confirmed, deceased, recovered \\ \hline
Netherland & Province & Rijksinstituut voor Volksgezondheid en Milieu Ministerie van Volksgezondheid, Welzijn en Sport &  deceased and recovered reported at country level \\ \hline
South Africa & Province & Nation Institute for Communicable Diseases  & confirmed, deceased, recovered  \\ \hline
Russian Federation & Federal Subjects & Federal Service Human Well Being &  confirmed, deceased, recovered \\ \hline
Saudi Arabia & Governorates & King Abdullah Petroleum Studies and Research Center &  confirmed, deceased, recovered \\ \hline
Sweden & Landsting  & Folkhälsomyndigheten &  confirmed and recovered  \\ \hline
Mexico & State & Gobierno De Mexico & confirmed and deceased  \\ \hline
Peru & Departamentos & Ministerio de Salud  & confirmed and recovered    \\ \hline
Brazil & State/ Federal District & Ministério da Saúde &  confirmed, deceased, recovered  \\ \hline
Portugal & Região & Ministério da Saúde &  confirmed, deceased, recovered  \\ \hline
Pakistan & Province & Ministry of National Health Services & confirmed, deceased, recovered  \\ \hline
France & Région/ Régions d'outre-mer & Saute Publique France & confirmed, deceased, recovered   \\ \hline
Ecuador  & Region & Ministerio de Salud & confirmed, deceased, recovered   \\ \hline
Indonesia  & Provinsi & Graha Badan Nasional Penanggulangan Bencana &  confirmed, deceased, recovered  \\ \hline
Philippine  & Region & Department of Health Datadrop &  confirmed, deceased, recovered   \\ \hline 
Liechtenstein & Country & Zurich Statistisches Amt  & confirmed, deceased, recovered  \\ \hline
China & Province/ SARs & DXY tracker & confirmed, deceased, recovered  \\ \hline
United States & States  & 1point3acres tracker & confirmed, deceased, recovered    \\ \hline
Canada & States & 1point3acres  tracker & confirmed, deceased, recovered    \\ \hline
\end{tabular}
\end{table*}

\end{document}